\documentclass[%
 preprint,
 amsmath,amssymb,
 aps,
]{revtex4-1}

\usepackage{graphicx}
\usepackage{dcolumn}
\usepackage{bm}
\usepackage[utf8]{inputenc}
\usepackage{hyperref}
\usepackage{color}

\begin{document}


\title{Infrared chemical imaging through nondegenerate two-photon absorption in silicon-based cameras}

\author{David Knez$^1$, Adam M. Hanninen$^1$, Richard C. Prince$^2$, Eric O. Potma$^{1,2}$\email{epotmaf@uci.edu}, Dmitry A. Fishman$^1$\email{dmitryf@uci.edu}}
\affiliation{$^1$Department of Chemistry, University of California, Irvine, CA 92697, USA}
\affiliation{$^2$Department of Biomedical Engineering, University of California, Irvine, CA 92697, USA}

\date{\today}

\begin{abstract}
Chemical imaging based on mid-infrared (MIR) spectroscopic contrast is an important technique with a myriad of applications, including biomedical imaging and environmental monitoring. Current MIR cameras, however, lack in performance and are much less affordable compared to mature Si-based devices, which operate in the visible and near-infrared. Here we demonstrate fast MIR chemical imaging through non-degenerate two-photon absorption (NTA) in a standard Si-based charge-coupled device (CCD). We show that wide-field MIR images can be obtained at 100 ms exposure times using picosecond pulse energies of only a few fJ per pixel through NTA directly on the CCD chip. Because this on-chip approach does not rely on phase-matching, it is alignment-free and does not necessitate complex post-processing of the images. We emphasize the utility of this technique through chemically selective MIR imaging of polymers and biological samples, including MIR videos of moving targets, physical processes and live nematodes.
\end{abstract}

\maketitle


\section*{\label{sec:intro}Introduction}
Many fundamental molecular vibrations have energies in the mid-infrared (MIR) window - a wavelength region that stretches from about 2 $\mu$m to 10 $\mu$m. For this reason, the MIR range is of particular interest for spectroscopic imaging. The ability to generate images with chemical selectivity is of direct relevance to a myriad of fields, driving the implementation of MIR-based imaging for biomedical mapping of tissues~\cite{Bhargava2012,Wetzel1999,Wrobel2018}, inspection of industrial ceramics~\cite{Su2014}, stand-off detection of materials~\cite{Wang2005}, mineral sensing~\cite{Meer2012,Chen2015}, and environmental monitoring~\cite{Mintenig2017}, among others. 

Given its unique analytical capabilities, it is perhaps surprising that MIR-based imaging is not a more widely adopted technology for chemical mapping. The relatively scarce implementation of MIR imaging has been due in part to the lack of bright and affordable light sources in this range, although recent developments in MIR light source technology have largely overcome this problem~\cite{Yeh2015,Borondics2018,Kilgus2018}. Nonetheless, a remaining limitation is the performance and high cost of MIR cameras. Current cameras are based on low bandgap materials, such as HgCdTe (MCT) or InSb, which inherently suffer from thermally excited electronic noise~\cite{Rogalski2002}. Cryogenic cooling helps to suppress this noise but it renders the MIR camera a much less practical and affordable detector compared to mature Si-based detectors for the visible and near-IR. 

Recognizing the attractive features of Si-based cameras, several strategies have been developed that aim to convert information from the MIR into the visible/NIR range, thus making it possible to indirectly capture MIR signatures with a Si detector. A very recent development is the use of an entangled MIR/vis photon pair, which allows MIR ghost imaging through detecting visible photons on a Si-based camera~\cite{Lemos2014}. Another strategy accomplishes the MIR-to-visible conversion by using a nonlinear optical (NLO) response of the sample. Photothermal imaging, which probes the MIR induced changes in the sample with a secondary visible beam, is an example of this approach~\cite{Lee2011,Zhang2016,Samolis2019,Bai2019,Schnell2020}. Yet another technique uses a nonlinear optical crystal placed after the sample to up-convert MIR radiation with an additional pump beam through the process of sum-frequency generation (SFG)~\cite{Johnson2012,Tidemand2016,Junaid2018,Gu2010,Thew2008,Watson1990,Barh2019} The visible/NIR radiation produced can be efficiently registered with a high bandgap semiconductor detector. Elegant video-rate MIR up-conversion imaging has recently been accomplished with a Si-based camera at room temperature, offering an attractive alternative to imaging with MCT focal plane arrays~\cite{Junaid2019}. A possible downside of SFG up-conversion techniques is the requirement of phase-matching of the MIR radiation with the pump beam in the NLO medium. This implies crystal rotation to enable the multiple projections needed for capturing a single image and post-processing for each measured frame for image reconstruction.

An alternative to utilizing an optical nonlinear response of the sample (photothermal) or a dedicated conversion crystal for indirect MIR detection (SFG up-conversion) is the use of the NLO properties of the detector itself. In particular, the process of nondegenerate two-photon absorption (NTA) in wide bandgap semiconductor materials has been shown to permit detection of MIR radiation at room temperature with the help of an additional visible or NIR probe beam~\cite{Fishman2011,Hayat2008,Pattanaik2016,Pattanaik2015}. In NTA, the signal scales linearly with the MIR intensity with detection sensitivities that rival that of cooled MCT detectors~\cite{Fishman2011}. Compared to SFG-based up-conversion, NTA does not depend on phase-matching and avoids the need for an NLO crystal altogether, offering a much simpler detection strategy. Moreover, the nonlinear absorption coefficient drastically increases with the energy ratio of the interacting photons~\cite{Cirloganu2010,Cirloganu2011,Hutchings1992,Sheik-Bahae1992,Wherrett1984}, allowing detection over multiple spectral octaves. Although NTA has been shown to enable efficient MIR detection with single pixel detectors, its advantages have not yet been translated to imaging with efficient Si-based cameras. Here, we report rapid, chemically selective MIR imaging using NTA in a standard CCD camera at room temperature.

The nature of nonlinear absorption enhancement for direct-band semiconductors has been extensively studied and modeled with allowed-forbidden transitions between two parabolic bands~\cite{Hutchings1992,Sheik-Bahae1992,Wherrett1984,Sheik-Bahae1991} The nonlinear absorption coefficient $\alpha_2$ for photon energies $\hbar\omega_{p}$ and $\hbar\omega_{MIR}$ can be written as~\cite{Sheik-Bahae1991}:
\begin{eqnarray}
    \alpha_2(\omega_p,\omega_{MIR})&=&K\frac{\sqrt{E_p}}{n_pn_{MIR}E^3_g}F(x_p,x_{MIR})\\
    F&=&\frac{(x_p+x_{MIR}-1)^{3/2}}{2^7x_p(x_{MIR})^2}\left(\frac{1}{x_p}+\frac{1}{x_{MIR}}\right)^2, ~x_p=\frac{\hbar\omega_p}{E_g},~x_{MIR}=\frac{\hbar\omega_{MIR}}{E_g}\nonumber
\end{eqnarray}
where $E_p$ is the Kane energy parameter, $n_p$ and $n_{MIR}$ are refractive indices and $K$ is a material independent constant. The function $F$ accounts for the change in nonlinear absorption as the ratio between the pump and MIR photon energies is adjusted, with dramatic enhancements when the pump energy is tuned closer to the bandgap energy $E_g$. For an indirect bandgap semiconductor like Si, optical transitions can be understood as a nonlinear process that involves three interacting particles - two photons and a phonon. Several models have been considered to describe multi-photon absorption in Si, including earlier ``forbidden-forbidden'' models~\cite{Dinu2003}, and the more recently suggested ``allowed-forbidden'' and ``allowed-allowed'' pathways~\cite{Garcia2006}. The latter two models agree well with degenerate absorption experiments~\cite{Bristow2007}. For the case of NTA, experiments demonstrate enhancement behavior similar to those seen in direct-bandgap semiconductors`\cite{Zhang2015,Poulvellarie2018}, with the ``allowed-allowed'' pathways providing the best description~\cite{Cox2019}. Modest numbers of the acquired and derived nonlinear absorption coefficients of only a few cm/GW have classified Si as a rather inefficient material for NTA. For this reason, attempts to develop MIR detection strategies based on Si detectors have been scarce. In this work, we show that, despite previous concerns, detecting MIR radiation through NTA in silicon is not only feasible, but readily provides a very practical approach for MIR imaging with standard cameras.

\section*{\label{sec:results}Results}

\subsection{MIR detection with a Si photodiode}
\begin{figure}[ht]
    \centering
    \includegraphics[width=17cm]{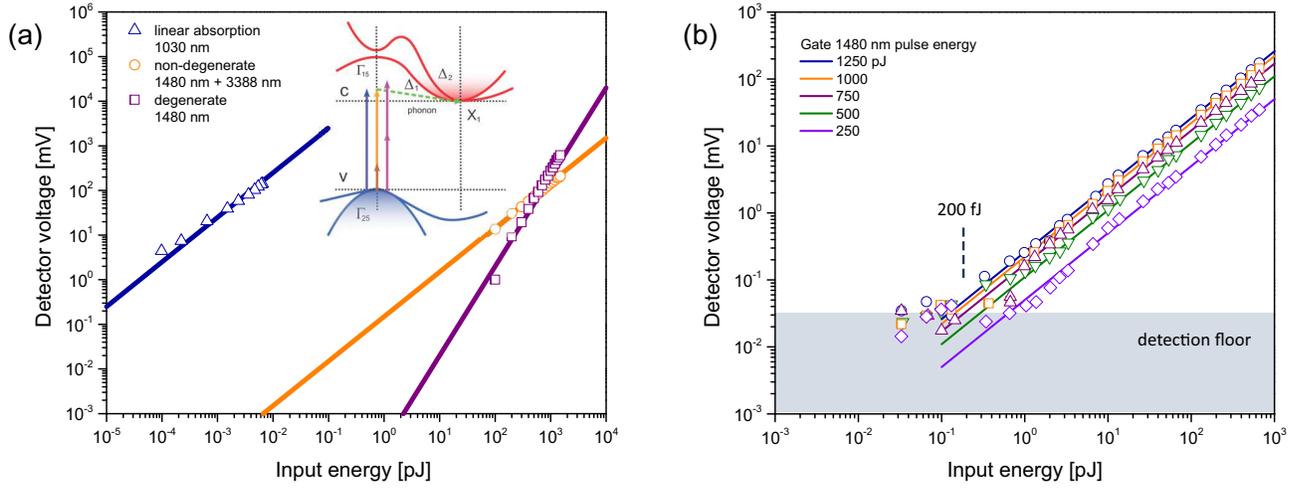}
    \caption{Detection of weak infrared radiation via non-degenerate two photon absorption in a Si photodiode. (a) Linear (blue) as a function of pulse energy at 1030~nm, non-degenerate (orange) as function of pump pulse energy at 1480~nm and degenerate (purple) as function of pump pulse energy at 1480~nm. For the non-degenerate curve, the MIR pulse energy at 3388~nm was set at 6.5~nJ. Inset: proposed scheme of photon absorption in Si. (b) Full dynamic range for MIR detection with detection floor of 200~fJ picosecond pulse energy for the given detector parameters.}
    \label{fig:fig1}
\end{figure}

We first discuss the utility of Si as a MIR NTA detector using picosecond pulses of low peak intensities. In Figure \ref{fig:fig1}a, we compare the linear absorption of 9708 cm$^{-1}$ (1030~nm) photons by a standard Si photodiode with that of NTA for a 2952 cm$^{-1}$ (3388~nm) MIR and 6756 cm$^{-1}$ (1480~nm) pump pulse pair. Since the 1030 nm photon energy exceeds the Si bandgap energy ($E_g\sim1.1$~eV (1100~nm)), strong one-photon absorption can be expected. Based on this measurement, the estimated responsivity is $R=0.2$~A/W, close to the reported response for Si detectors at 1030~nm. In the NTA experiment, the MIR and pump photon energies add up to the same energy (9708 cm$^{-1}$) as in the one-photon experiment, and we may thus expect a NTA response in Si, albeit weaker. The current photon energy ratio is $\hbar\omega_p$/$\hbar\omega_{MIR}=2.2$. The NTA response is shown in orange and compared with the degenerate two-photon absorption of the pump pulse. As expected, the NTA signal scales linearly with the NIR pulse energy. Previously reported values of $\alpha_{2d}\sim2$~cm/GW~\cite{Bristow2007} for degenerate and $\alpha_{2n}\sim5$~cm/GW~\cite{Cox2019} for non-degenerate cases with comparable photon ratio agree well with our observations. Note that there is a regime where the NTA is stronger than the degenerate two-photon absorption of the pump, using 6.5~nJ MIR pulse at 3388~nm.

We next study the sensitivity of MIR detection through NTA in Si. In Figure \ref{fig:fig1}b, the detected NTA signal is plotted as a function of the MIR pulse energy (at 2952 cm$^{-1}$) for various energies of the pump pulse. For these experiments, especially at higher NIR peak intensities, the degenerate contribution of pump pulse has been subtracted using modulation of the MIR beam and lock-in detection. We observe that the signal scales linearly with the MIR pulse energy for all settings. The minimally detectable MIR picosecond pulse energy is $\sim200$~fJ using rather modest NIR pump peak intensities. In previous work with a direct large-bandgap GaN detector, a detection limit of 100~pJ has been reported, using femtosecond pulses and a photon energy ratio $>10$~\cite{Fishman2011}. Here we observe higher detection sensitivities in Si, while using picosecond pulses and a much lower photon energy ratio. Such high detection sensitivities are remarkable, and are due in part to the favorable pulse repetition rate (76~MHz) used in the current experiment, offering much better sampling compared to kHz pulse repetition rates used previously. The strategy used here offers superior sensitivity, detecting 4 orders of magnitude smaller MIR peak intensities of 20~W/cm$^2$ (with 0.09~MW/cm$^2$ at 1480~nm pump pulse) versus 0.2~MW/cm$^2$ (with 1.9~GW/cm$^2$ at 390~nm pump pulse) as previously reported~\cite{Fishman2011}. Given that enhancement scales with the photon energy ratio, we may expect even greater sensitivities for experiments with higher pump photon energies and lower MIR photon energies, with a projected detection floor as low as a few tens of fJ (1~W/cm$^2$).

\subsection{MIR spectroscopy with a single pixel Si detector}
\begin{figure}[h]
    \centering
    \includegraphics[width=9cm]{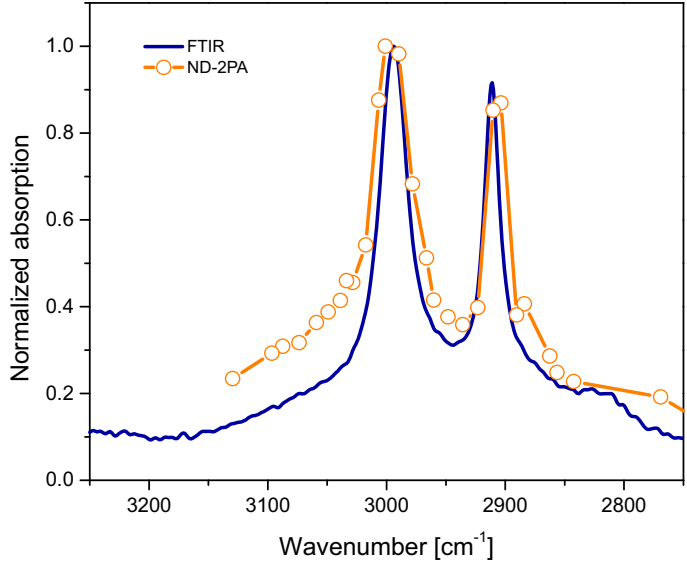}
    \caption{Absorption spectrum of dimethyl sulfoxide (DMSO) using non-degenerate two photon detection for measuring the transmitted MIR radiation. Results are in excellent agreement with spectrum measured obtained with conventional ATR-FTIR of bulk DMSO.}
    \label{fig:fig2}
\end{figure}

As an example of the utility of MIR detection with a Si photodetector, we perform a simple MIR absorption spectroscopy experiment by a 10 $\mu$m film of dimethyl sulfoxide (DMSO). For this purpose, we spectrally scan the MIR energy in the 2750--3150~cm$^{-1}$ range and detect the transmission MIR via NTA on the Si photodiode. The absorption of the film of this thickness is estimated as $\mbox{OD}=0.03$. The spectral resolution is determined by the spectral width of the picosecond pulse, which corresponds to $\sim5$~cm$^{-1}$. For these experiments, the MIR pulse has been kept at an average power of 15~mW ($\sim10$~kW/cm$^2$ peak intensity) while the NIR pump beam is set to 100~mW (66~kW/cm$^2$). Because the pump and MIR pulses are parametrically generated by the same source, there is no temporal walk-off on the picosecond timescale while performing the scan. The resulting DMSO absorption spectrum shows the characteristic lines associated with the symmetric and asymmetric C-H stretching modes43, which corroborates the Fourier transform IR (FTIR) absorption spectrum (Figure \ref{fig:fig2}, see Methods).

\subsection{MIR imaging through on-chip nondegenerate two-photon absorption in a CCD camera}
\begin{figure}[h]
    \centering
    \includegraphics[width=14cm]{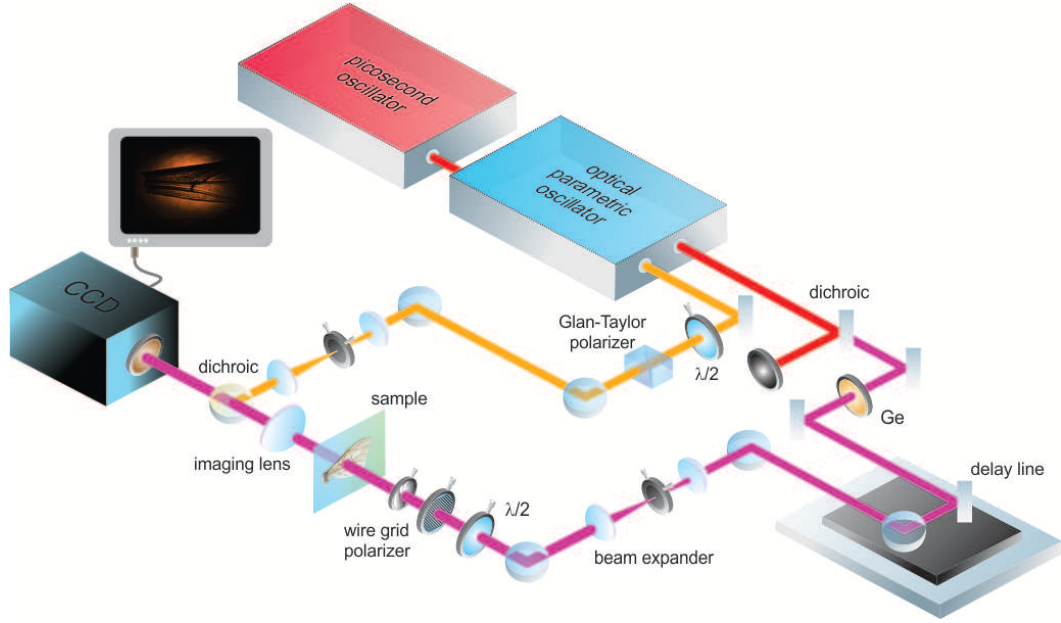}
    \caption{Schematic of wide-field MIR imaging system based on non-degenerate two-photon absorption in a Si-based CCD camera. }
    \label{fig:fig3}
\end{figure}

Given the excellent NTA performance of a single pixel Si detector, we next explore the feasibility of MIR imaging through direct on-chip NTA in a Si-based CCD camera. Figure \ref{fig:fig3} provides a schematic representation of the MIR wide-field imaging system based on NTA. The pump and MIR beams are generated by a 4~ps, 76~MHz optical parametric oscillator (OPO), and are expanded to a beam diameter of $\sim3$~mm. The MIR arm contains the sample and a 100~mm CaF$_2$ lens to map the image in a 1:1 fashion onto the CCD sensor. The pump beam is spatially and temporally overlapped with the MIR beam with the aid of a dichroic mirror, so that both beams are coincident on the CCD chip. Note that phase-matching is not important for NTA, implying that the angle of incidence of the pump beam can be adjusted freely. Here we use a conventional, Peltier-cooled CCD camera (Clara, Andor), featuring 6.45~$\mu\mbox{m}^2$ pixels in a $1392\times1040$ array. The current magnification and effective numerical aperture of the imaging lens ($NA=0.015$) provides an image with $\sim100~\mu$m resolution, corresponding to about 20 pixels on the camera. Though not the ultimate goal of current experiments, better spatial resolution can be easily achieved using focusing systems of higher numerical aperture.

\begin{figure}[ht]
    \centering
    \includegraphics[width=10cm]{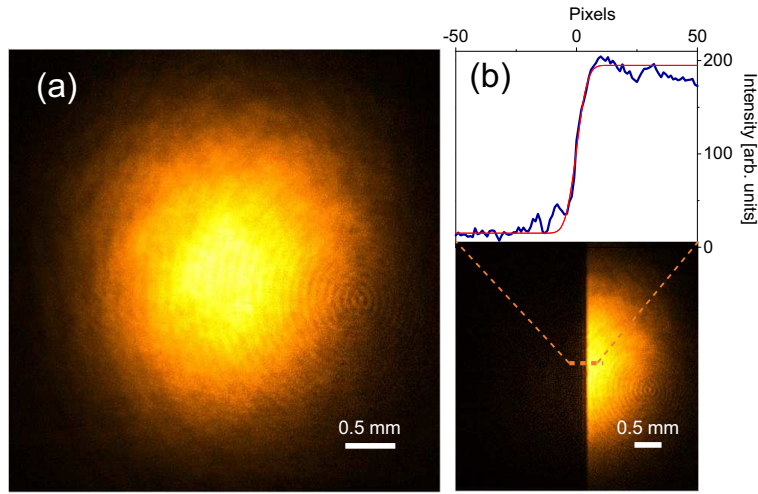}
    \caption{(a) Image of MIR (3394 nm) beam profile using a 1478~nm pump pulse.  (b) Image of razor blade covering half of MIR. The cross section is shown in the top of the panel. Error function analysis shows that the resolution is about 15 pixels ($\sim100~\mu$m) under the current conditions.}
    \label{fig:fig4}
\end{figure}

\begin{figure}[ht]
    \centering
    \includegraphics[width=14cm]{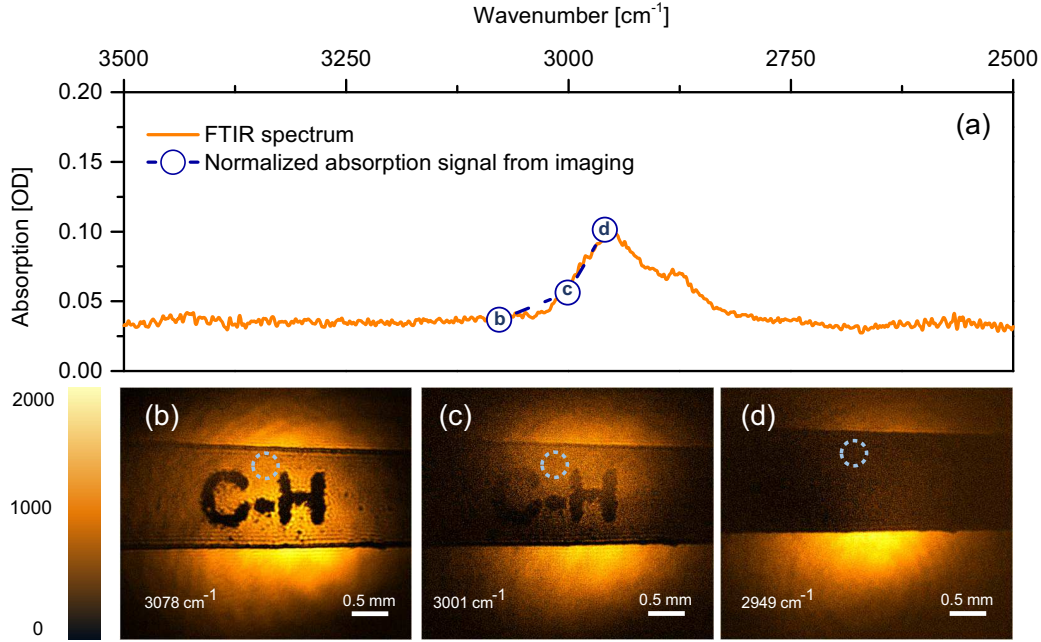}
    \caption{Spectral imaging of a 150~$\mu$m thick cellulose acetate film. The printed letters serve as a mask that blocks broadband radiation. (a) FTIR transmission spectrum. MIR image taken at (b) an off-resonance energy, (c) the high energy side of the absorption maximum and (d) the absorption maximum.}
    \label{fig:fig5}
\end{figure}

In Figure \ref{fig:fig4}a, we show the NTA image of the MIR beam projected onto the CCD sensor using 1~s exposure time. The degenerate background signal has been subtracted to solely reveal the MIR contribution. With the current experimental arrangement, background has to be measured only once for a given NIR pump intensity, and can be subtracted automatically during imaging, requiring no further post-processing. Here we have used peak intensities of $\sim1.5$~kW/cm$^2$ for the MIR beam and $\sim1.4$~kW/cm$^2$ for the NIR pump beam. Under these conditions, each camera pixel only receives pulse energies on the order of a few fJ. In Figure \ref{fig:fig4}b, we show the same MIR beam with a razor blade blocking one half of the beam, clearly emphasizing the attained MIR contrast. The fringing at the blade’s interface is a direct consequence of light diffraction at the step edge. More images of test targets, including MIR images of a USAF test chart, are provided in the \href{https://www.chem.uci.edu/~dmitryf/images/SI.pdf}{{\color{blue}Supplementary Information}} (Figure S2).
For the current experimental conditions, we find that MIR intensity changes on the order of 10$^{-2}$ OD in the image are easily discernible even with exposure times shorter than 1~s. To demonstrate chemical imaging capabilities, we perform MIR imaging on a $\sim150$~$\mu$m thick cellulose acetate sheet commonly used as transparencies for laser jet printing. Figure \ref{fig:fig5}a depicts the FTIR spectrum of cellulose acetate in the 2500--3500~cm$^{-1}$ range, showing a clear spectral feature due to C-H stretch vibrational modes. In Figure \ref{fig:fig5}b, a strip of the cellulose acetate sheet is imaged at 3078 cm$^{-1}$, off-resonant with the C-H stretching vibration. Transmission through the sheet is high because of the lack of absorption. To highlight the contrast, the letters ``C-H'' have been printed with black ink directly onto the material, providing a mask with limited transmission throughout the 2500--3500~cm$^{-1}$ range. When tuning into the CH-mode resonance (Figure \ref{fig:fig5}b, 3001~$^{-1}$), the transmission is seen to reduce, resulting in lower contrast between the ink and the film. When the MIR is tuned to the maximum of the absorption line (Figure \ref{fig:fig5}, 2949 cm$^{-1}$), the limited transmission through the film completely eliminates the ink/film contrast. The relative magnitude of MIR absorption, extracted from the images, maps directly onto the absorption spectrum in Figure \ref{fig:fig5}a, demonstrating quantitative imaging capabilities. The observed contrast is based on a rather modest absorption difference of only $7\times10^{-2}$ OD. More examples of MIR imaging of printed cellulose acetate samples can be found in the \href{https://www.chem.uci.edu/~dmitryf/images/SI.pdf}{{\color{blue}Supplementary Information}} (Figure S3).

\begin{figure}[ht]
    \centering
    \includegraphics[width=16cm]{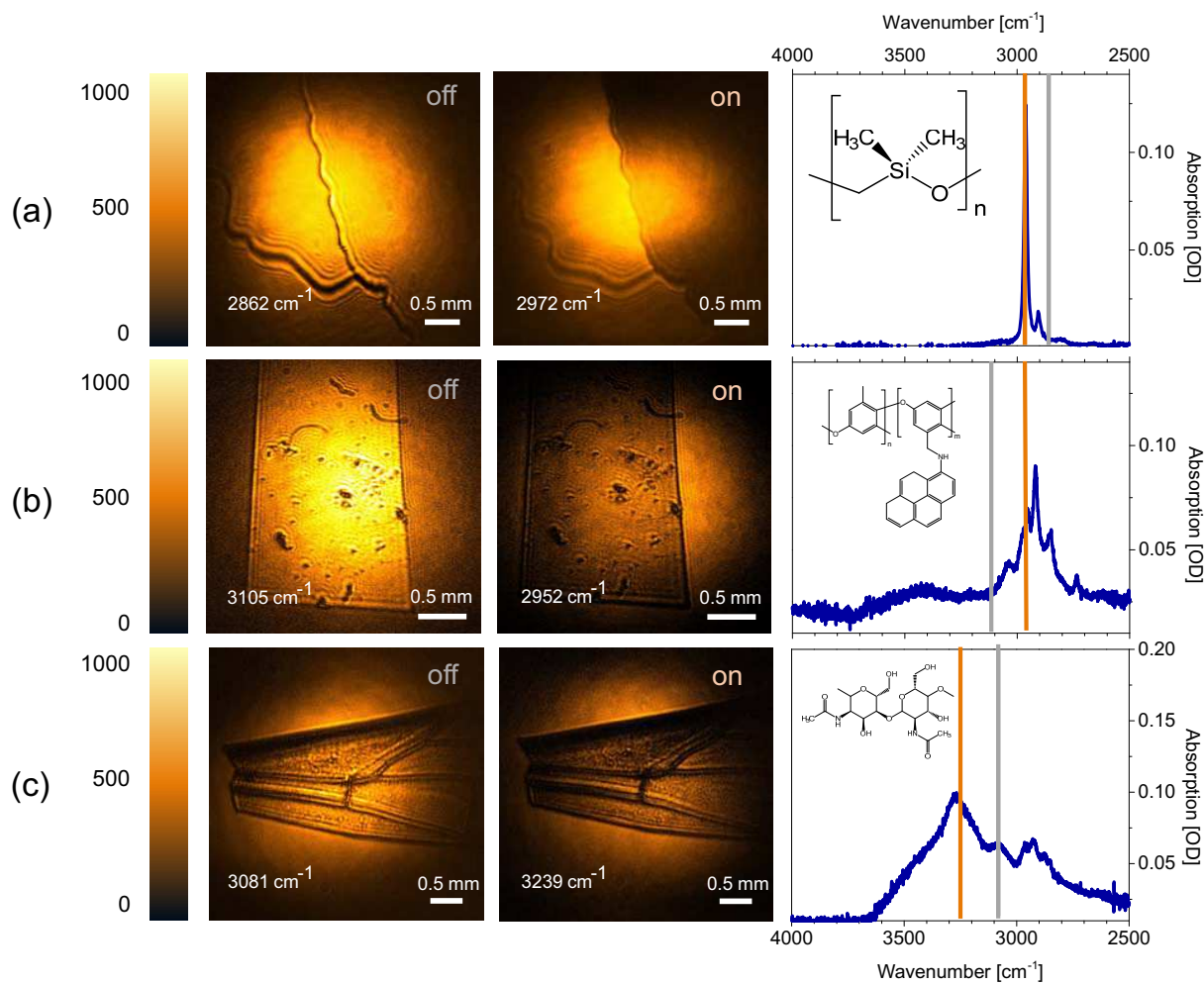}
    \caption{MIR images of various materials accompanied with corresponding FTIR spectra. Left column shows off resonance MIR images, whereas middle column shows MIR images taken at an energy that corresponds with a designated absorptive line. Right column displays the FTIR absorption spectra of the sample with on (orange) and off (grey) resonance frequencies indicated. (a) Interface between D$_2$O and silicone lubricant. (b) APPPO polymer film. (c) Wing of a common bee.}
    \label{fig:fig6}
\end{figure}

With the wide-field MIR imaging capabilities thus established, we highlight some examples of chemical imaging of several common materials. To suppress contrast due to refractive index differences, we have suspended the materials in (vibrationally non-resonant) D$_2$O to reveal true absorption contrast. Figure \ref{fig:fig6}a depicts the interface between D$_2$O and a $\sim20~\mu$m thick polydimethylsiloxane film, a silicon-based organic polymer commonly used as vacuum grease. The difference between the images taken on and off resonance with the methyl stretching mode reveals clear chemical contrast. Similarly, in Figure \ref{fig:fig6}b, chemical contrast is evident when tuning on and off the C-H stretching resonance of a $30~\mu$m membrane of poly(2,6-dimethylphenylene oxide-co-N-(2,6-dimethylphenylene oxide) aminopyrene), a material of considerable interest as an ion-exchange membrane. Lastly, we demonstrate MIR imaging of a bee’s wing in Figure \ref{fig:fig6}c, a rather complex natural structure that is rich in chitin. The chitin MIR spectrum in the 2500--3500~cm$^{-1}$ range contains overlapping contributions from OH-, NH- and CH-groups, resulting in broad spectral features. The absorption difference between the 3239~cm$^{-1}$ and 3081~cm$^{-1}$ vibrational energies is $\Delta$OD=0.04, yet the contrast difference is still evident from the NTA MIR image. 
\subsection{MIR videography of sample dynamics}
The signal strength is sufficient for MIR imaging at even faster acquisition rates. In the Supplementary Information, we provide MIR imaging through NTA with a 100 ms exposure time. Given that the current camera requires an additional 100 ms of readout time per frame, the effective imaging acquisition time was pushed to 5~fps. Under these conditions, we have recorded videos of several mechanical and physical processes as well as live microorganisms. First, in Video 1 (see \href{https://www.chem.uci.edu/~dmitryf/images/SI.pdf}{{\color{blue}Supplementary Information}}) the real-time movement of a printed target on cellulose acetate films is demonstrated, both under vibrationally off-resonance (\href{https://www.chem.uci.edu/~dmitryf/images/VideoV1a.avi}{{\color{blue}V1a}}) and resonance (\href{https://www.chem.uci.edu/~dmitryf/images/VideoV1b.avi}{{\color{blue}V1b}}) conditions. \href{https://www.chem.uci.edu/~dmitryf/images/VideoV2.avi}{{\color{blue}Video 2}} shows a live recording of the dynamics of the immersion oil droplet placed atop the CaF$_2$ window, under vibrationally resonant conditions. The flowing droplet can be seen with clear chemical contrast in real time. Moreover, one can observe the formation of intensity fringes near the edge of droplet due to Fresnel diffraction and interference within the oil film, i.e. Newton's ring effect.
In \href{https://www.chem.uci.edu/~dmitryf/images/VideoV3.avi}{{\color{blue}Video V3}}, we show NTA-based MIR imaging of live nematodes suspended in D$_2$O buffer, recorded at 3381~nm (2958 cm$^{-1}$). Image contrast is due to absorption by the methyl stretching vibrations of protein, in addition to refractive effects. The video demonstrates that active, live nematodes can be captured in real time under the MIR illumination conditions used in NTA detection. 

\section*{Discussion}
In this work, we have shown that the principle of NTA can be extended to MIR imaging by direct on-chip two-photon absorption in a CCD camera. This principle enabled us to acquire images at 100 ms exposure times at fJ-level picosecond pulse energies per pixel, experimental conditions that allows wide-field MIR imaging of live, freely suspended organisms at reasonably high frame rates. The use of a CCD camera serves as an attractive alternative to standard MIR cameras, such as cryogenically cooled MCT detectors. NTA enables good quality MIR images without cryogenic cooling, significantly reducing the complexity and costs of the detector. In addition, NTA-enabled imaging benefits from the mature technology in Si-based cameras, offering robust and affordable detection solutions. These advantages are not at the expense of sensitivity, as previous work based on MIR femtosecond pulses has shown that NTA offers comparable sensitivity to (single pixel) MCT detectors~\cite{Fishman2011}. 

Unlike other recent methods for improving MIR detection with visible/NIR detectors, our method takes MIR light as its direct input. Photothermal imaging, for instance, relies on MIR-induced optical changes in the sample (expansion, refractive changes), which are subsequently probed with a vis/NIR beam. An alternative method but related method is acoustic detection of the MIR photothermal effect, which has recently been demonstrated~\cite{Shi2019}. NTA-based imaging is independent of secondary effects due to MIR sample illumination, as it registers intensity changes in the MIR directly. Similarly, our new NTA imaging approach differs fundamentally from SFG up-conversion techniques. While the latter also uses a visible/NIR camera to generate MIR-based images, the SFG up-conversion mechanism is based on a separate step that converts the MIR light with the help of a pump beam into visible radiation in an external nonlinear optical crystal~\cite{Barh2019}. To enable high conversion efficiencies, specific orientations of the crystal are needed so that a portion of the MIR image is phase-matched in the SFG process. To capture a full image, a rapid sampling of crystal orientations must be applied, followed by an image reconstruction step. Although impressive MIR imaging capabilities have been achieved with the SFG up-conversion technique, the NTA method approaches comparable imaging performance while using MIR and pump intensities that are an order of magnitude lower. NTA avoids the external light conversion step, and thus significantly simplifies the overall imaging system. Because NTA does not rely on phase-matching, it can generate images in a single shot and forgoes the need for post-acquisition image reconstruction. 

Although the current work shows the feasibility of NTA-based MIR imaging, the approach can be further improved to achieve even better performance. For instance, we have used an off-the-shelf CCD camera that is not optimized for long wavelength applications. The sensor is protected with a 1.5~mm thick silica window that shows significant MIR attenuation of an estimated OD=1. It is trivial to replace the window with a sapphire flat, which would boost the MIR transmission by at least an order of magnitude. The frame rate of our current camera is limited, and higher frame rates can be easily achieved with more advanced models. In addition, higher NTA efficiencies can be obtained with shorter pulses. The use of high-repetition rate fs pulses would allow imaging at much lower average powers while maintaining high efficiency. It should be noted that detector arrays based on materials other than Si are also interesting for NTA applications. GaAs, for instance, exhibits significantly higher two-photon absorption efficiencies and a steeper band edge absorption than Si, which are both favorable for MIR detection through NTA. Finally, the practical implementation of the NTA imaging technique requires the availability of a pump beam in addition to a MIR source. Although OPO systems constitute a natural choice because of their broadly tunable synchronized pump/idler pulse pairs, recent developments in MIR light source technology promise alternative and more compact solutions, including efficient frequency conversion with long wavelength fiber lasers. Such developments will likely improve the practical implementation of NTA-based imaging for a wide range of applications.

\begin{acknowledgments}
We thank Prof. Shane Ardo, Leanna Schulte and Cassidy Feltenberger for use of the APPPO film shown in Figure 6b. We thank Prof. Joseph Patterson for help with nematode sample handling. This work was funded by the National Institutes of Health, R01GM132506. We thank the Laser Spectroscopy Labs for help with linear spectroscopy characterizations.
\end{acknowledgments}

\section*{Methods}

\subsection{FTIR experiments} Conventional infrared absorption spectra were measured using a Jasco 4700 FTIR spectrometer both in transmission and attenuated total reflection (ATR) geometries. For ATR experiments we used the Jasco ATR-Pro One accessory equipped with a diamond crystal. The spectra were averaged over 20 scans and acquired with a 2 cm$^{-1}$ resolution, close to the resolution of the corresponding picosecond NTA experiments.

\subsection{Sample handling} Most of the prepared samples were suspended in D$_2$O to suppress refractive effects and thus reveal pure absorption contrast. DMSO and silicone lubricant (Dow Corning) were obtained from Sigma Aldrich and used without further purification. The sample materials, including the APPPO polymer film and clipped bee wings, were immersed in D$_2$O and confined between hermetically sealed 1~mm thick CaF$_2$ windows (diameter = $1''$). Experiments on cellulose acetate films were performed in air without the use of CaF$_2$ windows. C. elegans were obtained from Carolina Biological. Nematodes were picked up from agar plates with filter paper, immersed in a phosphate buffered saline D$_2$O solution and placed between two CaF$_2$ windows spaced by a 50~$\mu$m Teflon spacer. 

\subsection{Non-degenerate two photon absorption detection with Si photodiode} We used a conventional Si photodiode (FDS100, Thorlabs) with parameters described in \href{https://www.chem.uci.edu/~dmitryf/images/SI.pdf}{{\color{blue}Supplementary Information}}. The transparent window in front of the Si material was removed to improve MIR transmission. The experiments were performed in pump-probe geometry with the setup depicted in Figure \ref{fig:fig3}, without utilizing a separate imaging lens in the MIR arm. Both MIR and NIR beams were focused onto the Si photodiode by a $f=100$~mm broadband achromat  (Trestle Optics)~\cite{Hanninen2019}. NIR intensity was varied by combination of half-wavelength plate and Glan-Thompson polarizer. The MIR intensities were controlled by another half-waveplate and a wire-grid polarizer. The polarization of both NIR and MIR optical pulses were linear and parallel, and were kept constant throughout the experiments. The MIR beam was modulated at 160~Hz by a mechanical chopper and the NTA signal contribution was isolated by a lock-in amplifier (SR510, Stanford Instruments).

\subsection{Imaging using CCD camera} We used a Si-based CCD camera (DR-328G-CO2-SIL Clara, Andor). The setup is explained in Figure 3 of the main text. We used a 1:1 imaging system with a $f=100$~mm CaF$_2$ lens to project the image onto the CCD camera (Clara, Andor). For live nematode imaging the imaging system was changed to a 2x magnification, using a 4f imaging system composed of a $f=50$~mm and a $f=100$~mm broadband achromat (Trestle Optics)~\cite{Hanninen2019}.
\clearpage
\pagebreak


\end{document}